\documentclass[sigconf,screen]{acmart}

\AtBeginDocument{%
  \providecommand\BibTeX{{%
    \normalfont B\kern-0.5em{\scshape i\kern-0.25em b}\kern-0.8em\TeX}}}

\setcopyright{acmcopyright}
\copyrightyear{2021}
\acmYear{2021}
\acmPrice{15.00}
\acmDOI{10.1145/3411764.3445767}
\acmISBN{978-1-4503-8096-6/21/05}
\acmConference[CHI '21]{CHI Conference on Human Factors in Computing Systems}{May 8--13, 2021}{Yokohama, Japan}
\acmBooktitle{CHI Conference on Human Factors in Computing Systems (CHI '21), May 8--13, 2021, Yokohama, Japan}

\acmSubmissionID{7721}

\begin{document}

\title[Patterns for Representing Situational Robotic Knowledge Graphs]{Patterns for Representing Knowledge Graphs to Communicate Situational Knowledge of Service Robots}

\author{Shengchen Zhang}
\email{shengchenzhang1207@gmail.com}
\affiliation{%
  \institution{College of Design and Innovation, Tongji University}
  \streetaddress{281 Fuxin Rd}
  \city{Shanghai}
  \country{China}
}

\author{Zixuan Wang}
\email{zixuanwang@tongji.edu.cn}
\affiliation{%
  \institution{College of Design and Innovation, Tongji University}
  \streetaddress{281 Fuxin Rd}
  \city{Shanghai}
  \country{China}
}

\author{Chaoran Chen}
\email{chenchaoran@tongji.edu.cn}
\affiliation{%
  \institution{College of Design and Innovation, Tongji University}
  \streetaddress{281 Fuxin Rd}
  \city{Shanghai}
  \country{China}
}

\author{Yi Dai}
\email{amyyue0416@163.com}
\affiliation{%
  \institution{College of Design and Innovation, Tongji University}
  \streetaddress{281 Fuxin Rd}
  \city{Shanghai}
  \country{China}
}

\author{Lyumanshan Ye}
\email{2623706775@qq.com}
\affiliation{%
  \institution{College of Design and Innovation, Tongji University}
  \streetaddress{281 Fuxin Rd}
  \city{Shanghai}
  \country{China}
}

\author{Xiaohua Sun}
\email{xsun@tongji.edu.cn}
\affiliation{%
  \institution{College of Design and Innovation, Tongji University}
  \streetaddress{281 Fuxin Rd}
  \city{Shanghai}
  \country{China}
}

\renewcommand{\shortauthors}{Zhang and Wang, et al.}

\begin{abstract}
  Service robots are envisioned to be adaptive to their working environment based on situational knowledge. Recent research focused on designing visual representation of knowledge graphs for expert users. However, how to generate an understandable interface for non-expert users remains to be explored.
  In this paper, we use knowledge graphs (KGs) as a common ground for knowledge exchange and develop a pattern library for designing KG interfaces for non-expert users. 
  After identifying the types of robotic situational knowledge from the literature, we present a formative study in which participants used cards to communicate the knowledge for given scenarios. We iteratively coded the results and identified patterns for representing various types of situational knowledge.
  To derive design recommendations for applying the patterns, we prototyped a lab service robot and conducted Wizard-of-Oz testing. The patterns and recommendations could provide useful guidance in designing knowledge-exchange interfaces for robots.
\end{abstract}

\ccsdesc[500]{Human-centered computing~Graphical user interfaces}
\ccsdesc[500]{Human-centered computing~User interface design}
\ccsdesc[300]{Computer systems organization~Robotics}
\ccsdesc[500]{Computing methodologies~Cognitive robotics}
\ccsdesc[100]{Human-centered computing~User studies}


\keywords{Design Patterns, Interface Design, Knowledge Graph, Human-Robot Interaction}


\begin{teaserfigure}
  \includegraphics[width=\textwidth]{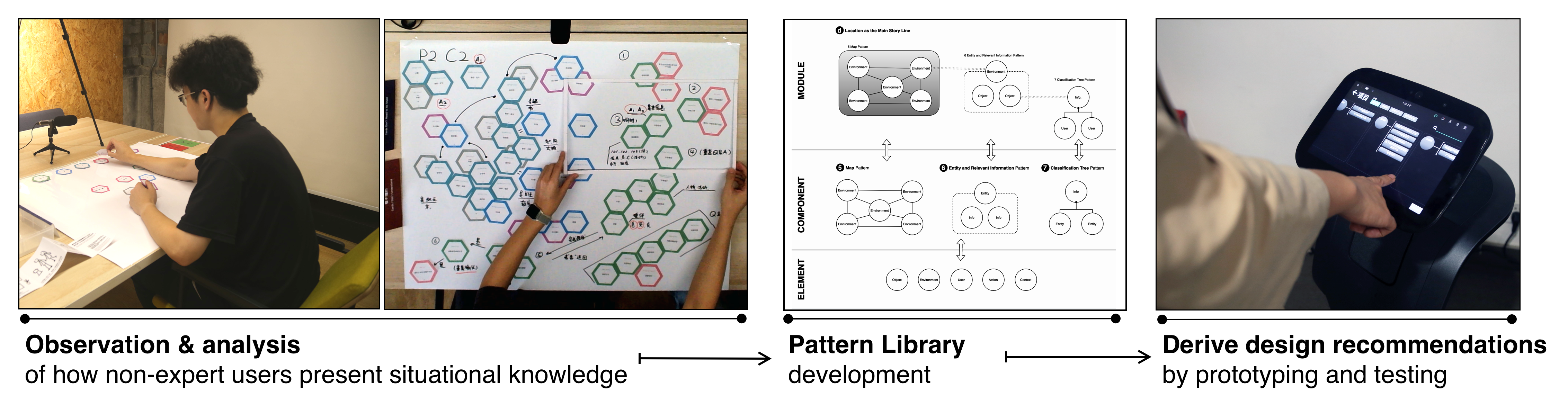}
  \caption{In this work, we use observations and analysis of human presentation of situational knowledge to develop a pattern library for interface design. We further derive design recommendations by prototyping and testing the interface of a service robot.}
  \Description{A three-step image illustrating the process of our research framework. We use observations and analysis of human presentation of situational knowledge to develop an interface design pattern library. We further derive design recommendations by prototyping and testing the interface of a service robot.}
  \label{fig:teaser}
\end{teaserfigure}

\maketitle

\section{Introduction}

Service robots are increasingly pervasive in our daily life and are anticipated to execute complex tasks based on high-level goals and interact with users in an easily-understandable way\cite{kattepur2019roboplanner}. This requires robots to effectively organize and represent situational knowledge --- the in-situ information about humans, objects, places, and events in the robot's working environment\cite{javia2016knowledge}. 
However, since situational knowledge is often highly related to users and objects on the spot, robots need to interact with humans to mediate the mismatch between perception and comprehension of situational knowledge\cite{fang2015embodied}. 

Knowledge graphs\cite{hogan2020knowledge} are a common method used in both artificial intelligence (AI) to incorporate human knowledge into AI models, and in robotics to build knowledge-enabled robots. 
Previous research have been exploring designing interfaces to help domain experts to directly view and manipulate the knowledge graph in a robot in order to understand and operate it\cite{lemaignan2017artificial}. 
However, for non-expert users who have little technical knowledge in ontological models like KGs, it can be hard to understand and interact with such interfaces. Existing works are developing a set of human-friendly vocabulary to build robot ontology\cite{diprose2012people}, but research on KG interface design for non-expert users is still relatively few.

Meanwhile, previous works showed that design patterns benefit the field of HCI and HRI by providing reusable solutions for recurring problems\cite{kahn2008design}. For KG interfaces, the study of interface design patterns is especially useful. A set of patterns can lower the requirement on the understanding of technical details for the designers, while also accelerating the development of such interfaces.

This paper describes the process (Figure \ref{fig:teaser}) of developing a library of patterns for visually presenting situational, robotic knowledge graphs. The goal of the patterns is to help designing effective interfaces to communicate the situational knowledge of service robots. 

Our contribution is three-fold:
\begin{itemize}
    \item We developed a pattern library of high-level and low-level patterns of situational knowledge communication that can be used to design robot interfaces. 
    \item We described in detail our process of constructing the library from how non-expert users present KG elements on a canvas.
    \item We further derived design recommendations for using the pattern library through prototyping and a Wizard-of-Oz field study.
\end{itemize}

\section{related work}
Our work builds on prior research on situational knowledge exchange in humans and robots, challenges of knowledge graph in cognitive robot, and patterns of human robot interaction.

\subsection{Situational knowledge exchange in humans and robots}

Situational knowledge exchange has played an increasingly critical role in human-robot interaction, for both robotic perception and comprehension\cite{javia2016knowledge}. Many studies have focused on leverage symbol grounding techniques (connecting numeric and symbolic representation of real-world objects and the environment) to make situational knowledge exchange more accurate and easier to understand \cite{bastianelli2013knowledge}. These techniques have been used to support situational knowledge representation for robots\cite{topp2017interaction,pronobis2012large,nieto2010semantic}, situation assessment in human-robot interaction\cite{riley2010situation,sisbot2011situation}, and context-based human-robot collaboration\cite{zachary2013context,zachary2015context}.

Most relevant to our work is the research focused on grounding and representing situational knowledge for robots in a human-understandable way. For example, Bastianelli et al.\cite{bastianelli2013knowledge} proposed using metric maps and a multi-model interface that allowed users to guide robots to ground symbolic information in the operational environment. Carpenter et al.\cite{carpenter2017using} presented CARIL, a computational architecture that used declarative situational representation to adapt robots' behavior to humans during their collaboration. While several aspects of exchanging situational knowledge have been studied, such as creating robot task plans\cite{paxton2018evaluating,kattepur2019roboplanner} and programming the Robotic Operating System (ROS)\cite{tiddi2017ontology,leonardi2019trigger,diprose2012people}, few studies explored a unified form of visual representation for situational knowledge in robots.

Knowledge graphs can represent heterogeneous and interconnected knowledge\cite{jiang2020improving}, therefore often used as a unified format of knowledge representation in robots. Our work aims to identify the patterns and represent situational knowledge based on the knowledge graph in a service robot.

\subsection{Challenges of knowledge graph in cognitive robots}
Although knowledge graphs have been applied to facilitate human-robot interaction, it is hindered by three main challenges: static systems, monotonous interaction modality and intricate interfaces.

Firstly, robots based on static systems are hard to adapt to user preferences and intuitively interact with users \cite{jokinen2019human,javia2016knowledge}. To address this problem and explore human-friendly systems, some recent research enhanced the accuracy of human-robot knowledge communication by acquiring multi-modal human behavior data or the AR interface \cite{kennington2017graphical,liu2018interactive,tsiakas2017interactive}.

Secondly, previous research has indicated that uni-modal interaction (typically voice interaction for robots) might lead to misunderstanding the users' requests\cite{kennington2017graphical}. Liu et al. proposed a graphic interface that adds synchronous visual feedback of users' decision sequence based on a dialogue interaction system, which successfully reduced misunderstandings\cite{liu2010ambiguities}. It demonstrates the effectiveness of GUI-based interaction in human-robot knowledge communication. 

Thirdly, there is still relatively few research focused on enhancing non-expert users' understanding of ontological models and developing interactive systems for this purpose. Some relevant research explored human-readable ways of describing robot behavior\cite{diprose2012people}, and designed an interface using ontological abstraction to help non-experts users to program robots with low learning cost\cite{tiddi2018user}. These works inspired us to explore how to generate a systemic and understandable interface for exchanging knowledge between non-expert users and service robots.

\subsection{Patterns of human robot interaction}

Pattern language was first proposed by Alexander in 1977\cite{alexander1977pattern}, and then thrived in the field of Human computer interaction(HCI)\cite{pea1987user,seffah2010evolution} and human robot interaction(HRI)\cite{kahn_design_2008,mioch2014interaction,sauppe2014design}.
Kahn et al. argued that design patterns could benefit HRI by providing designers with the necessary knowledge and save their time when reusing these patterns to solve recurring problems\cite{kahn2008design}. Patterns also serve as scaffolding to help future research explore their relevant direction.

Prior research focused on extracting patterns to guide the design of HRI. For example, Oliveira et al. yielded social interaction patterns by observing human and robot players' behaviors in card games\cite{oliveira2018friends}. Oguz et al. developed an ontological framework from human interaction demonstrations, which could be transferred to HRI scenarios\cite{oguz2019ontology}.

Most relevant to our research are the patterns of presenting robot knowledge on an graphical interface. Robots have shown the ability to use interface to effectively communicate various types of information and promote task efficiency in teleoperating\cite{barba2020tele,chen2007human},manufacturing\cite{marvel2020towards} and elderly care\cite{klakegg2017designing}. Although prior research showed interest in exploring the design of interfaces, few focused on extracting patterns of knowledge presentation on the robot's interface. Our work aims to facilitate more design of robot interface to exchange situational knowledge.

\section{Formative Study}

Alexander points out in his foundational work\cite{alexander_timeless_1979} that patterns ``cannot be made, but only be generated, indirectly, by the ordinary actions of the people.''
Regarding applying design patterns to HRI, Freier et al. further argued that design patterns in nature are ``patterns of human interaction with the physical and social world\cite{kahn_design_2008}.''
In line with this reasoning, we grounded our formative study in observations and analysis of non-expert users' methods to visually communicate situational knowledge. Participants are tasked with presenting the knowledge needed in an HRI scenario using cards on a canvas. We then iteratively coded the results and identified patterns for presenting various types of robotic situational knowledge. The following sections describe the process of discovering and formalizing these patterns.

\subsection {Scenarios}
\label{section:scenarios}

To create a comprehensive and believable setting to ground our observation, we first identify three HRI scenarios that span three categories of situational knowledge exchange interactions: semantic, procedural and episodic. As knowledge communication involves both robot and human parties, we referenced research in both human knowledge \cite{knowledgetypes,types_qualities_knowledge} and robotic knowledge\cite{laird_standard_2017}. We adopted the categorization in \cite{laird_standard_2017} as the knowledge types are applicable to both humans and robots, and are more robot-specific.

\paragraph{Communicating semantic knowledge} Semantic knowledge is "semantically abstract facts\cite{laird_standard_2017}." 
In semantic scenarios, the participants are asked to present information about user information, object ownership, or environment locations on the canvas.

\paragraph{Communicating procedural knowledge} Procedural knowledge is "knowledge about actions, whether internal or external\cite{laird_standard_2017}." 
In procedural scenarios, the participants are asked to present information about robot tasks or actions in a planned procedure.

\paragraph{Communicating episodic knowledge} Episodic knowledge is "contextualized experiential knowledge"\cite{laird_standard_2017}. 
In episodic scenarios, the participants present knowledge such as a perceived sequence of events or information related to a certain period of time.

To ensure our results' applicability to a wide variety of working environments, we expanded each scenario into three types of contexts from private to public: home, office, and elderly care center. This results in nine scenarios used in the formative study.

We use comic strips to present each scenario to the participants to reduce narration's influence on the results. An example of the scenario comic is shown in Figure~\ref{fig:scenario-comic}. The complete collection of comics can be found in the supplementary materials.

\begin{figure}[h]
  \centering
  \includegraphics[width=\linewidth]{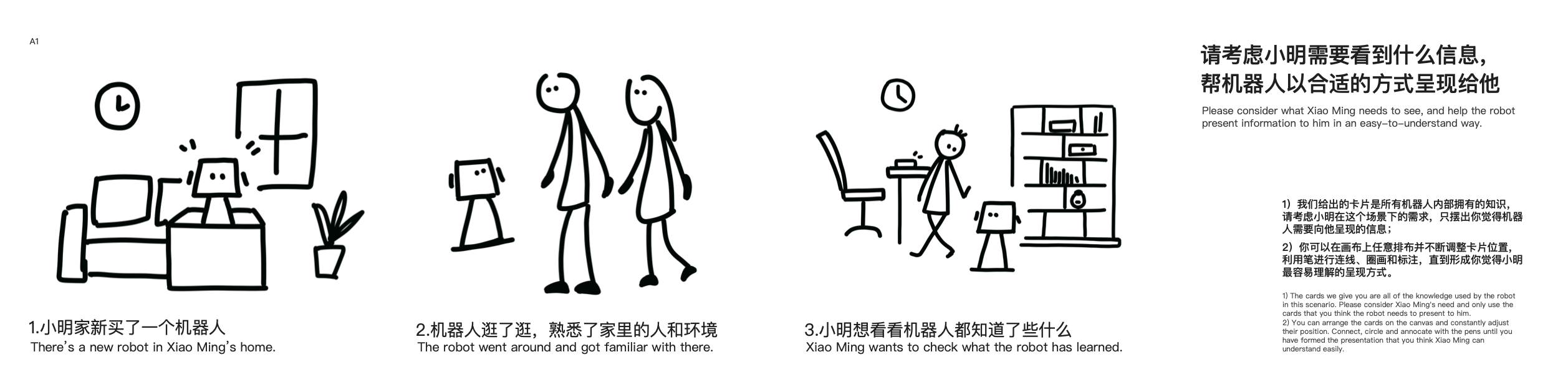}
  \caption{An example of the scenario comic}
  \Description{An example of the scenario comic. It consists of four panels. The first three depict a scene where a user asks the robot what it has learned on its first day at home. The fourth panel has instructions for the participant. It asks the participant to play the role of the robot, and arrange cards on the canvas to display the information.}
  \label{fig:scenario-comic}
\end{figure}

\begin{figure}[h!]
  \centering
  \includegraphics[width=\linewidth]{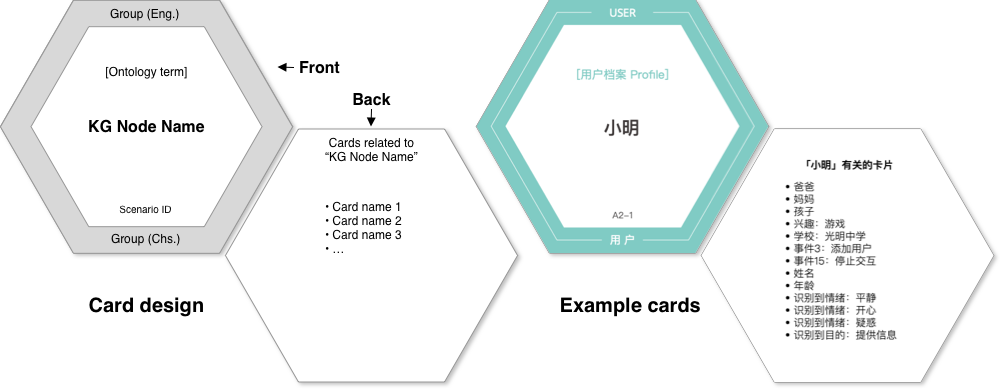}
  \caption{The design of the knowledge cards. On the right is an example of a card in the group ``User'', with ontology term ``User profile''. }
  \Description{The design of the knowledge cards. The card is hexagonal with shaded outlines. The ontology group that the card belongs to is written in the shaded area, near the top and bottom of the card. In the center is the name of the KG node and its ontology term. There is also a small text-box showing the scenario ID that the card belongs to. The back of the card features a list of related card names. On the right of the image is an example of a card in the group ``User'', with ontology term ``User profile''. }
  \label{fig:card-design}
\end{figure}

\subsection{Knowledge Cards} 

To achieve a more comprehensive coverage of the types of situational knowledge in service robots, we conducted a literature survey of works on ontology for service robots in the Scopus database\footnote{https://www.scopus.com}. Intially, we used ``situational'' and its variations (such as ``situated'' and ``in situ'') in a pilot search. To the best of our efforts, we found no existing works categorizing robot situational knowledge types. Therefore, we searched for works categorizing general knowledge types of robot, using the query shown below:

\begin{verbatim}
    {robot AND knowledge graph} OR {robot AND 
    ontology} OR {robot AND semantic network} OR 
    {robot AND knowledge model} AND NOT industr* 
    AND NOT agricultur* AND NOT farming
\end{verbatim}

The search returned 228 papers. We further excluded works that focused on a specific type of knowledge and those that focused on specific applications of robot ontology. In the end, 11 papers were selected. We reviewed each paper for ontology terms and grouped similar terms to form an initial list of robot knowledge types. We finally removed non-situational knowledge about robotic components and capabilities to arrive at Table~\ref{tab:knowledgetype}.

\begin{table*}
  \caption{List of situational knowledge used in our formative study}
  \label{tab:knowledgetype}
  \begin{tabular}{llp{2.8in}p{1.4in}}
    \toprule
    Group       & Ontology              & Description & Examples\\
    \midrule
    Objects     & Object
                & 
                A physical entity in the environment, excluding agents such as users and robots.
                & \cite{olivares-alarcos_review_2019}
                \cite{lim_ontology-based_2011}
                \cite{il_hong_suh_ontology-based_2007}
                \cite{chang2020ontology}
                \cite{jeon_ontology-based_nodate}
                \cite{tenorth_knowrob_2013}
                \cite{azevedo_using_2019}
                \cite{lemaignan_artificial_2017}
                \cite{kruijff_situated_2007}
                \\ 
                
                & Affordance
                & 
                The possibility of action\cite{olivares-alarcos_review_2019} on an object, as perceived by the robot. 
                & \cite{olivares-alarcos_review_2019}
                \cite{jeon_ontology-based_nodate}
                \cite{tenorth_knowrob_2013}
                \\
    Environment & Environment Map       
                & 
                A spatial representation of the robot's working environment\cite{olivares-alarcos_review_2019}. 
                & \cite{olivares-alarcos_review_2019}
                \cite{lim_ontology-based_2011}
                \cite{il_hong_suh_ontology-based_2007}
                \cite{chang2020ontology}
                \cite{jeon_ontology-based_nodate}
                \cite{tenorth_knowrob_2013}
                \cite{lemaignan_artificial_2017}
                \cite{kruijff_situated_2007}
                \\
    Users
                & User profile          
                & 
                The basic information of the users\cite{chang2020ontology}, such as name and ID in the face recognition module.
                & 
                \cite{lemaignan_artificial_2017}
                \cite{chang2020ontology}
                \cite{jeon_ontology-based_nodate}
                \cite{tenorth_knowrob_2013}
                \cite{azevedo_using_2019}
                \cite{mahieu_semantics-based_2019}
                \cite{bruno_knowledge_2019}
                
                \\
                & Social concept        
                & 
                Social knowledge of the user obtained while interacting with a human.\cite{chang2020ontology}
                & \cite{chang2020ontology}
                \cite{mahieu_semantics-based_2019}
                \cite{bruno_knowledge_2019}
                \\
                & Emotion/Intention     
                & 
                The emotional state and intention of the user, as detected by the robot.
                & \cite{chang2020ontology}
                \cite{azevedo_using_2019}
                \cite{mahieu_semantics-based_2019}
                \\
    Action      & Action \& Task       
                & 
                A task represents ``a piece of work that has to be done by the robot\cite{olivares-alarcos_review_2019}''. An action refers to ``a way to execute a task\cite{olivares-alarcos_review_2019}'', which we use to denote a step that the robot takes to complete a task. 
                &
                \cite{olivares-alarcos_review_2019}
                \cite{lim_ontology-based_2011}
                \cite{il_hong_suh_ontology-based_2007}
                \cite{tenorth_knowrob_2013}
                \cite{mahieu_semantics-based_2019}
                \cite{lemaignan_artificial_2017}\\
                
                & Activity \& Behavior 
                & 
                The behavior that the robot adopts when carrying out a task. Example behaviors may include scripted interactions, open dialogue, or simply command-and-response. 
                & \cite{olivares-alarcos_review_2019}
                \cite{lim_ontology-based_2011}
                \cite{il_hong_suh_ontology-based_2007}
                \cite{chang2020ontology}
                \cite{jeon_ontology-based_nodate}
                \cite{tenorth_knowrob_2013}
                \\
                
                & Plan \& Method       
                & 
                A sequence of actions that the robots would take in order to fulfill a task. This can be either programmed or generated.
                & \cite{olivares-alarcos_review_2019}
                \cite{chang2020ontology}
                \cite{tenorth_knowrob_2013}
                \\
                
                & Interaction \& Communication 
                & 
                A group of pre-defined actions that involves interacting or communicating with the user. This type of knowledge is often associated with one or more types of behavior.
                & \cite{olivares-alarcos_review_2019}
                  \cite{chang2020ontology}
                  \cite{tenorth_knowrob_2013}
                  \cite{kruijff_situated_2007}
                \\
                
    Context     & Spatial \& temporal context 
                & 
                \cite{lim_ontology-based_2011} used these terms to refer to the spatial and temporal relationship between objects. We expand this definition to include non-object concepts such as events and interactions.
                & \cite{lim_ontology-based_2011}
                  \cite{lemaignan_artificial_2017}
                \\
                
                & Situation 
                & 
                \cite{lim_ontology-based_2011} used this terms to refer to the detected spatial status of objects (such as ``crowded''). We expand this definition to general social situations. 
                & \cite{lim_ontology-based_2011}
                  \cite{lemaignan_artificial_2017}
                \\
                
                & Event 
                & 
                An event is a notable happening detected by the robot.
                & \cite{tenorth_knowrob_2013}
                \\
    
  \bottomrule
\end{tabular}
\end{table*}

We then manually compiled knowledge graph datasets for the nine scenarios, with each dataset spanning all types of knowledge listed. The total number of nodes is kept roughly the same for each dataset to maintain similar task difficulty across the scenarios. 

We used hexagonal cards to represent each node in the knowledge graph, inspired by Padilla et al. \cite{padilla_understanding_2017}, as hexagon allows for versatile arrangement and efficient use of space\footnote{The source files we used to print the cards can be found at \url{https://github.com/tongji-cdi/robot-knowledge-canvases/tree/master/Cards\%20PDF}.}. 
An example of the card design is shown in Figure~\ref{fig:card-design}.

\subsection{Set up}
\begin{figure}[h!]
  \centering
  \includegraphics[width=\linewidth]{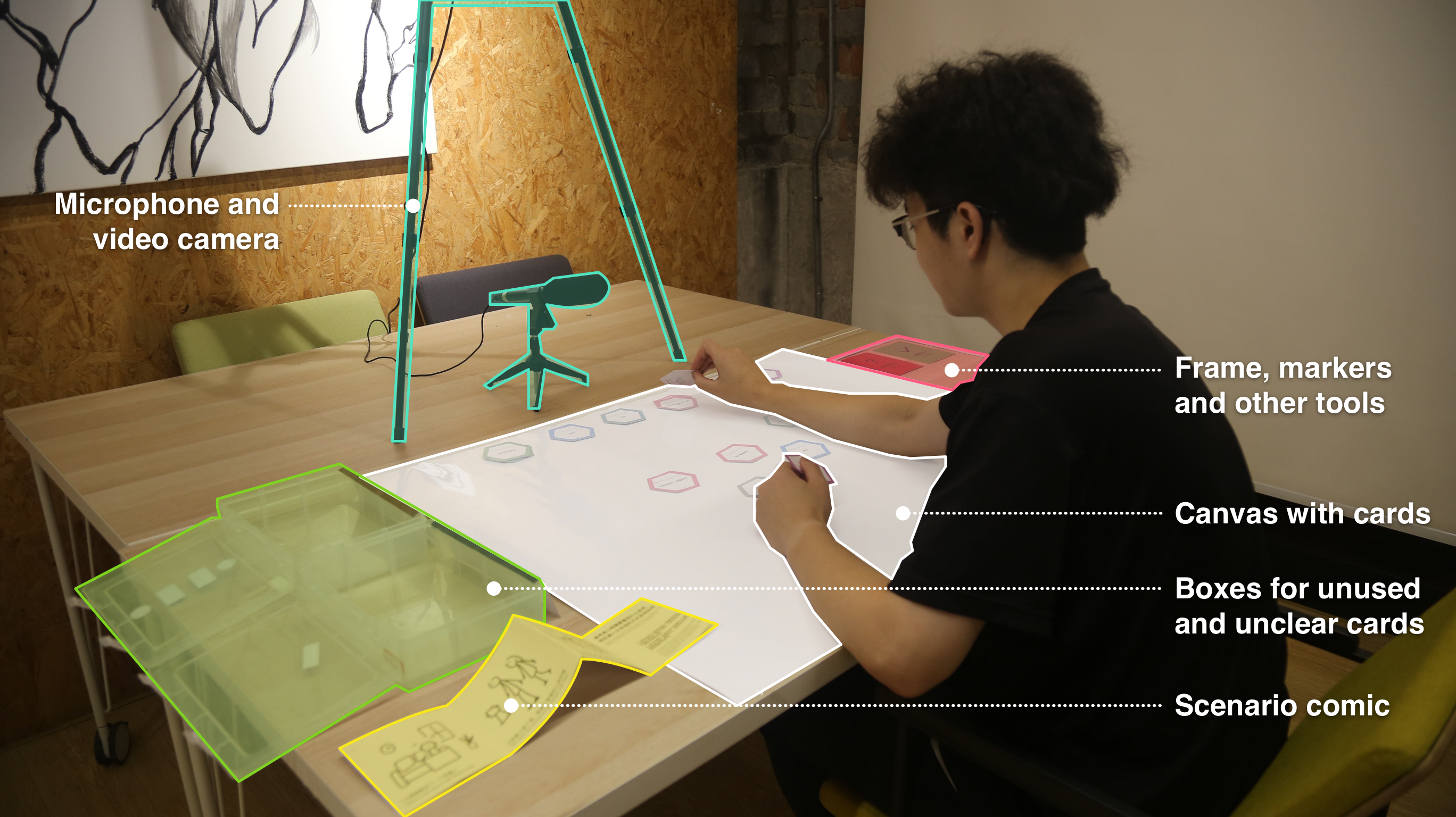}
  \caption{A photo of our study set-up. A researcher plays the role of the participant to protect the participants' identity.}
  \Description{An image showing our study set-up. Here, a researcher plays the role of the participant for anonymity reasons. The image shows a tabletop. On the tabletop there are several items. In front of the participant is a large canvas with six stacks of cards on it. The participant is taking some cards from the stacks and arranging them on the canvas. On the left, there are two boxes for unused and unclear cards. A scenario comic strip is also there. On the right, there's a rectangular frame, along with markers and other tools. Farther away from the participant is a microphone and a video camera on a tripod. The camera is out of the picture, and points down at the canvas.}
  \label{fig:Set-up Photo}
\end{figure}

The study was carried out in a controlled environment. As shown in Figure~\ref{fig:Set-up Photo}, the set up consisted of a table with a sheet of erasable canvas and hexagonal cards sorted by knowledge types. We provided markers and erasers for annotation. There are also two empty boxes for participants to put unused and unclear cards.

\subsection{Procedure/Task}
After signing a consent form, the participant can browse the material and ask questions. We then asked the participant to arrange the knowledge cards on the canvas. They are to arrange the cards in a way that they think best communicates the information as required in the scenario comic. The researcher leaves the working area during the process and observes the participant through a live camera feed. The participant can use the "question" and "finished" cards to signal the researcher when needed. There's no time limit to complete the task.

\begin{figure}[h]
  \centering
  \includegraphics[width=\linewidth]{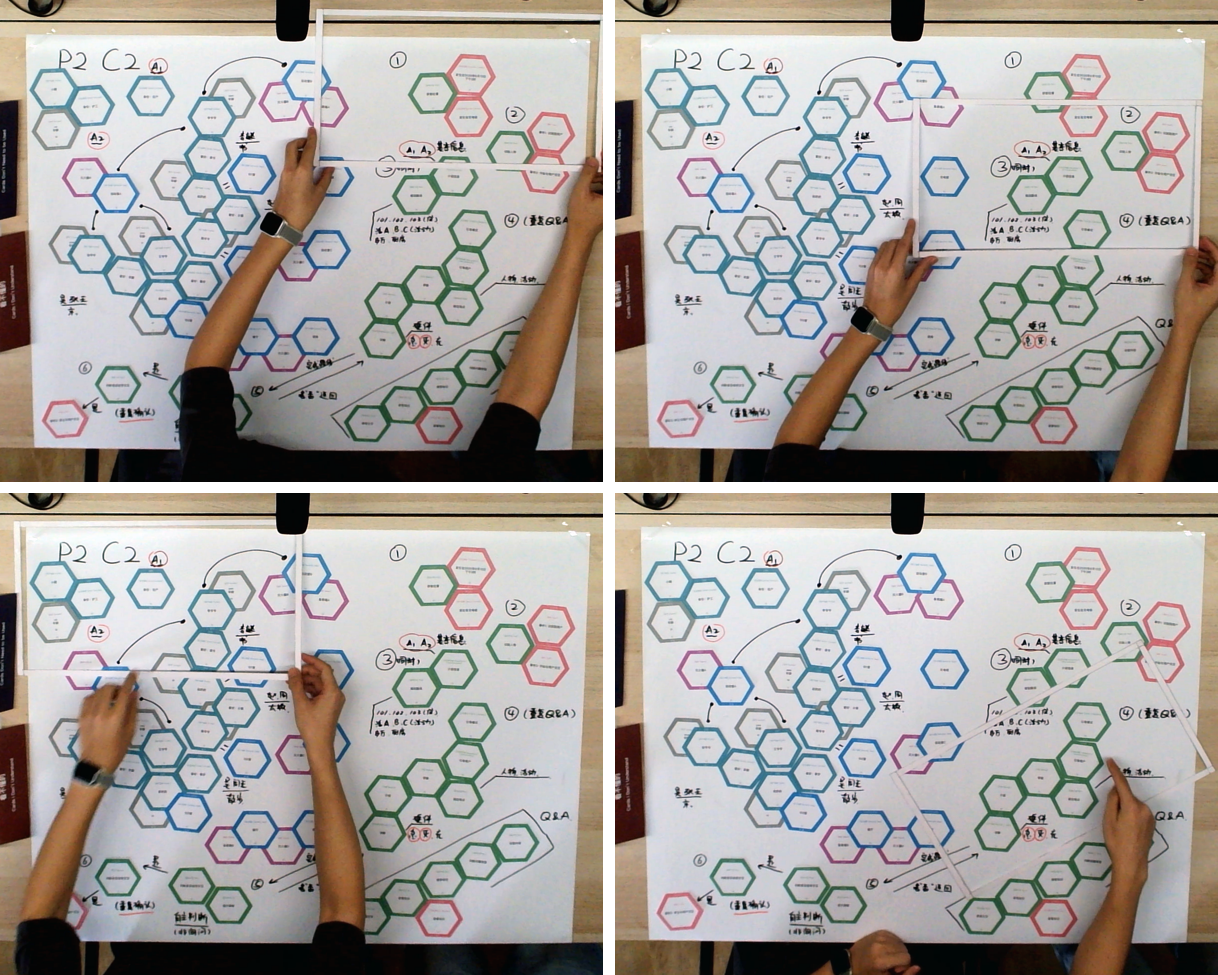}
  \caption{An example of the narration process of one of the participants (P2).}
  \Description{An example of the narration process of  participant P2 for scenario C2. The participant is moving a rectangular frame across the canvas, stopping here and there to point to and narrate the cards inside the frame.}
  \label{fig:formative-narration}
\end{figure}

After the creation, the participant is asked to communicate the knowledge on the display to the researcher using a frame. Figure~\ref{fig:formative-narration} shows an example of the narration. By defining the task in this manner, it allowed us to identify the grouping and sequence of the cards presented. Finally, the researcher interviews the participant about the arrangement logic and other interesting phenomena noticed during the process. 

In total, each participant would go through three similar sessions with scenarios communicating different knowledge types (semantic, procedural, episodic), which were randomly paired with three working environments (home, office, elderly care center). After that, we will ask participants to compare the knowledge representations they did with the three finished canvases shown on a screen.

\subsection{Data Collection}
A total of 12 university students aged from 21 to 27 (M = 23.83, SD = 1.51) from diverse fields took part in this study. We exclude participants knowledgeable in KG or graph visualization to avoid interference from their prior knowledge. 

During the study, a camera was used to record the behavior of the participants. We only recorded the canvas area to maintain as much anonymity as possible. A photo of the canvas was taken after each session to record the result\footnote{The canvases created by the participants can be found at \url{https://github.com/tongji-cdi/robot-knowledge-canvases}.}. All interviews were recorded.

\subsection{Data Analysis}

\begin{figure*}[h!]
  \centering
  \includegraphics[width=\linewidth]{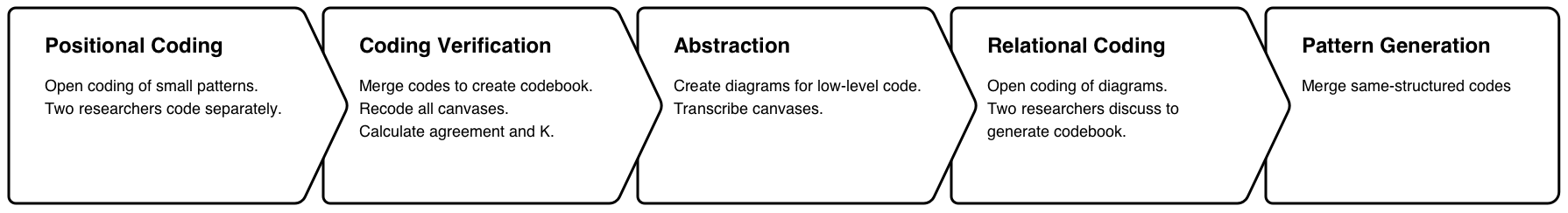}
  \caption{The process of our analysis.}
  \Description{A diagram showing the process of our analysis. The steps are the same as described in Section 3.6.}
  \label{fig:analysis-method}
\end{figure*}

Grounded theory is a common method for qualitative pattern generation in content and thematic analysis. However, it is mostly used to analyze interview text\cite{martin1986grounded}. Methods for analyzing diagram images and generating patterns from them are less explored. A relevant approach is visual grounded theory\cite{konecki2011visual}, which analyzes the results of visual narrations and iteratively codes data into various categories. Although it is often used to code images from realistic scenes, the work of Chapman et al.\cite{chapman2017picture} inspired us. We developed a process that consists of five steps: positional coding, coding verification, abstraction, relational coding, and pattern generation. An overview of our methodology can be seen in Figure~\ref{fig:analysis-method}.

\begin{figure*}[h!]
  \centering
  \includegraphics[width=\linewidth]{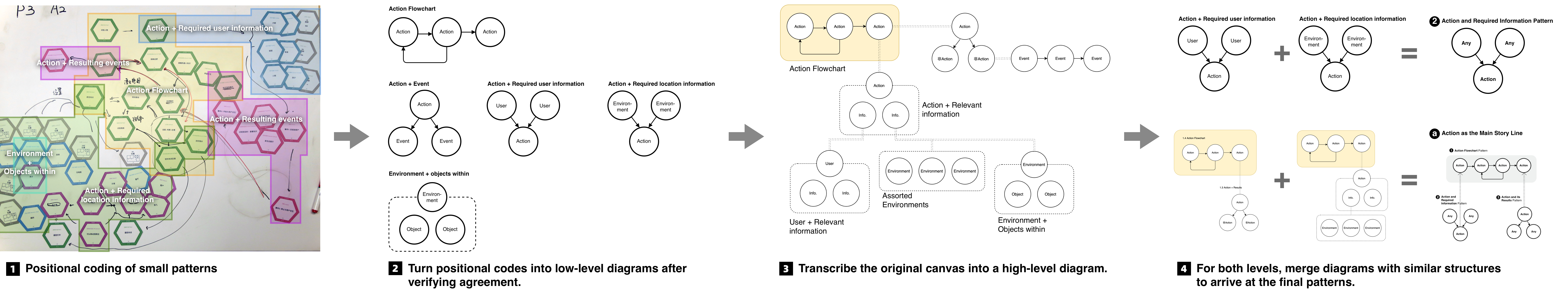}
  \caption{An illustration of the analysis process. }
  \Description{An illustration of the abstraction process. This diagram shows four stages. The first panel show small patches selected from the canvas image, and coded with different names. The second panel shows abstract representation of the small patterns using circles and arrows. The third panel shows that the content of the canvas image is substituted by abstract diagrams from the second panel. The last panel shows that similar diagrams are merged to become the final patterns.}
  \label{fig:anlaysis-transcription}
\end{figure*}

\paragraph{Positional coding: } Two researchers open coded any interesting phenomenon that emerged from the canvas photos separately. These phenomena are coded as nodes in NVivo\footnote{The NVivo source file is released under \url{https://github.com/tongji-cdi/robot-knowledge-canvases/releases/download/1.1/Coding_result.nvp}.}.
As a result of this step, each researcher generated a set of codes with description and examples. 

\paragraph{Coding verification: } The two researchers exchanged the set of codes and reviewed them together. The review process consisted of merging similar codes and standardizing naming conventions. This process results in a codebook that consists of code names and criteria for inclusion. After that, two researchers separately re-coded the images according to the codebook. Comparison of the two researchers' coding results shows reliable coding agreement (99\% agreement, Cohen's $\kappa$ = .79). 

\paragraph{Abstraction: } The researchers first established a standard graphical library of low-level codes and used them to transcribe the canvas photos into abstract diagrams. An example of this process can be seen in Figure~\ref{fig:anlaysis-transcription}. By doing so, we abstracted the relationship between the codes away from the cards' position and names. 

\paragraph{Relational coding: } One researcher open coded the diagrams generated from the previous step. During coding, the researcher focused on discovering patterns in the overall composition of low-level codes while paying particular attention to what constitutes the main storyline of the diagram. This coding process resulted in a set of high-level codes. A second researcher reviewed the coding results and discussed with the first until a consensus is reached. The two researchers then compiled the final codebook for high-level patterns.

\paragraph{Pattern generation: } The two researchers worked together to review the diagrams of low-level and high-level codes. They identified diagrams that share similar structures and merged the codes to arrive at the final patterns. In the end, we identified eight low-level patterns and four high-level patterns. The relations between these patterns are analyzed to form a pattern library.

\begin{figure}[h!]
  \centering
  \includegraphics[width=\linewidth]{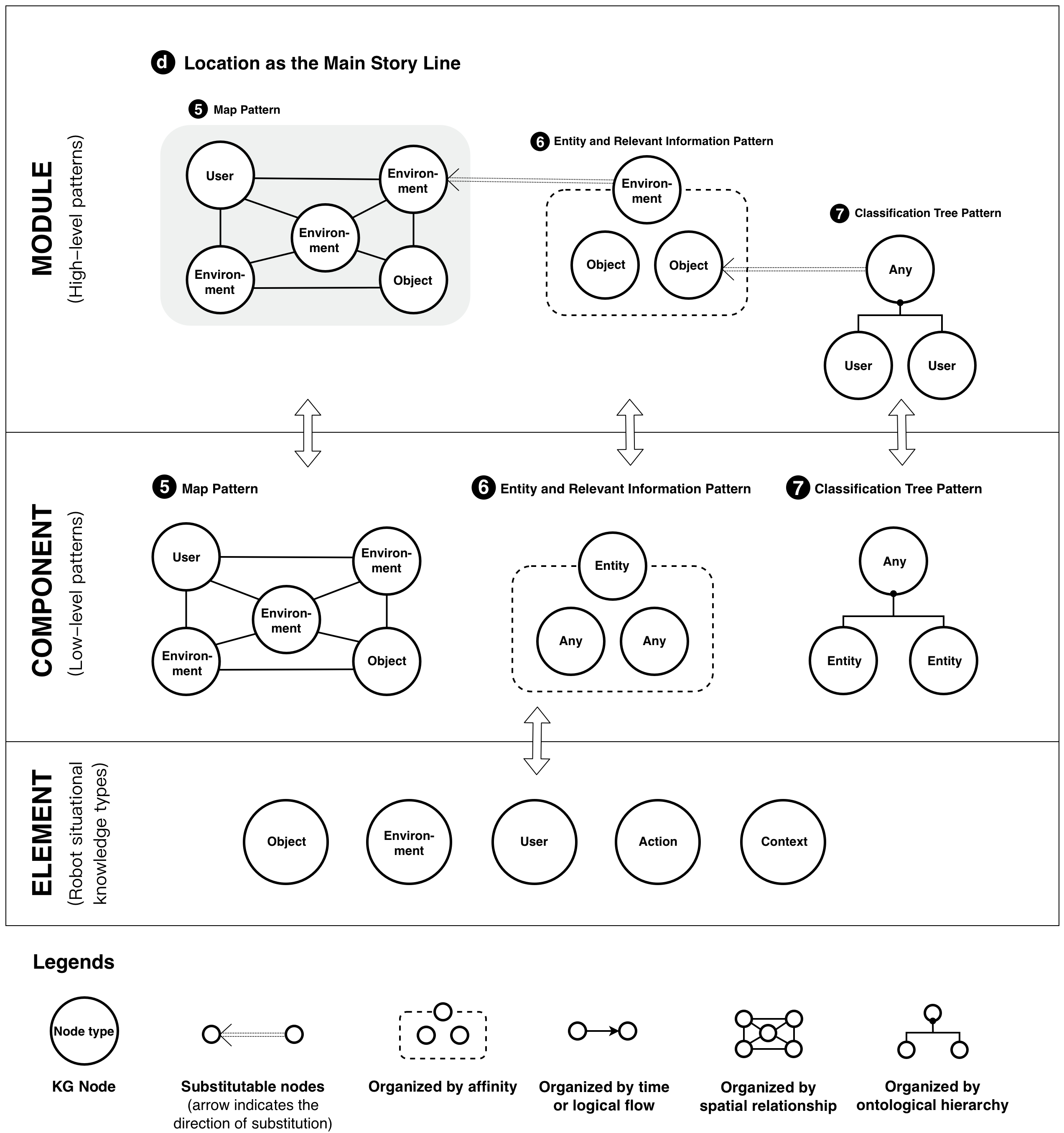}
  \caption{An excerpt of our pattern library. The component and module layer corresponds to the low-level and high-level patterns we discuss below. The elements corresponds to ontology groups defined in Table \ref{tab:knowledgetype}. The syntax described in the legends are used throughout this paper.}
  \Description{An excerpt of our pattern library. There are three layers: element, component, and module. The component and module layer corresponds to the low-level and high-level patterns we discuss below. The elements corresponds to ontology groups defined in Table 1.}
  \label{fig:pattern-library}
\end{figure}

\section{Pattern Library}

The pattern library we proposed consists of three parts: element, component, and module. The element part includes all types of the robotic situational knowledge we analyzed in Table~\ref{tab:knowledgetype}. The components (low-level patterns) are a typical combination and representation of certain elements. The modules (high-level patterns) are further built upon the components and can be easily applied to an interface. The structure of our pattern library can be seen in Figure~\ref{fig:pattern-library}, which is an example of Module d (Location as the Main Story Line).

\subsection{Low-level Patterns}
\begin{figure}[h]
  \centering
  \includegraphics[width=\linewidth]{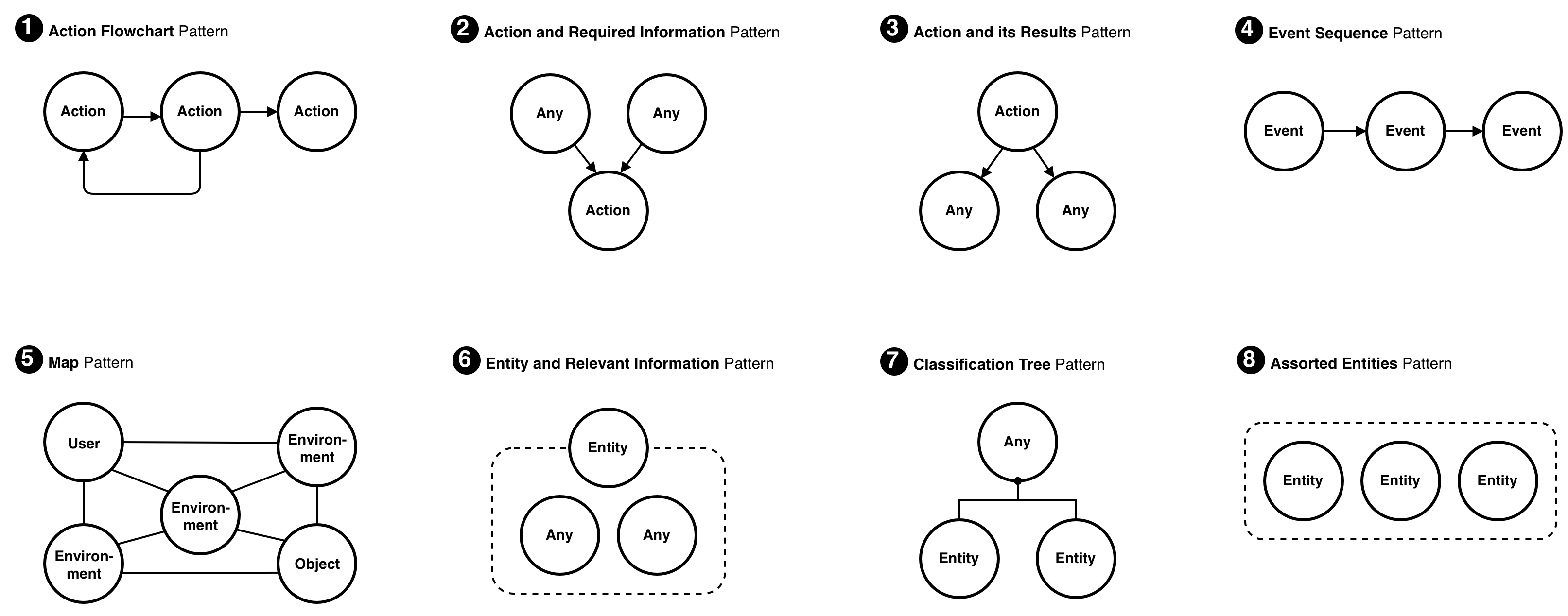}
  \caption{An overview of low-level patterns in our pattern library. ``Any'' in a diagram means a node of any type. ``Entity'' denotes something that physically exist, such as a user, an object, or a location.}
  \Description{An overview of low-level patterns in our pattern library. Eight patterns are present, each corresponding to a section below.}
  \label{fig:low-level-patterns}
\end{figure}

Low-level patterns are found as components that make up the whole canvas and mostly communicate a single message. These patterns can be employed when designing interface components that display specific information. An overview of low-level patterns can be found in Figure~\ref{fig:low-level-patterns}. The following sections provide a detailed description of these patterns.

\hypertarget{pattern-1}{}
\paragraph{1) Action Flowchart:}
    This pattern occurs when the participants want to show the logical flow of a series of actions. The inner structure connecting the actions in this pattern can be sequential, branching, and looping, similar to a flowchart.

\hypertarget{pattern-2}{}
\paragraph{2) Action and Required Information: }
    This pattern is used to show that the robot used pre-existing knowledge in order to undertake an action. It takes the shape of a central action card connected to its required information. The knowledge type of the required information can be any of the semantic types, including \textit{environment}, \textit{object}, and  \textit{user}. 

\hypertarget{pattern-3}{}
\paragraph{3) Action and Its Results: }
    This pattern is often similar in shape and content as \hyperlink{pattern-2}{\textit{Action and Required Information}}. However, it is used to show that an action of the robot resulted in the addition of certain new knowledge. A typical example is when the robot recognizes a new agent in its working place. Then it may add the knowledge automatically or by asking the user related.

\hypertarget{pattern-4}{}
\paragraph{4) Event Sequence: }
    This pattern is used to show a series of events that happened in chronological order. However, not all events are necessarily shown. Participants often skip events that they deem irrelevant. 
    
\hypertarget{pattern-5}{}
\paragraph{5) Map: }
    This pattern is used to showcase a collection of entities (e.g., environment, users, and objects) in their corresponding locations, similar to a real-world map. The pattern occurs when the knowledge communicated is highly related to locations, and there's no apparent order in the knowledge (e.g., time and alphabetical order).

\hypertarget{pattern-6}{}
\paragraph{6) Entity and Relevant Information: }
    This pattern is employed to show detailed information about the central entity, which can be any semantic knowledge. A typical example is when a \textit{user} card is placed. The participant will surround it with relevant information like age and room location. 

\hypertarget{pattern-7}{}
\paragraph{7) Classification Tree: } 
    This pattern is typically a tree-like structure, depicting the hierarchical classification of a collection of entities of the same type.
    An example is when a participant is trying to show all the people in an office. They used the department information to sort the people into two groups.

\hypertarget{pattern-8}{}
\paragraph{8) Assorted Entities: }
    This pattern describes a cluster of entities of the same type, without any specific order.

\subsection{High Level Patterns}

\begin{figure}[h!]
  \centering
  \includegraphics[width=\linewidth]{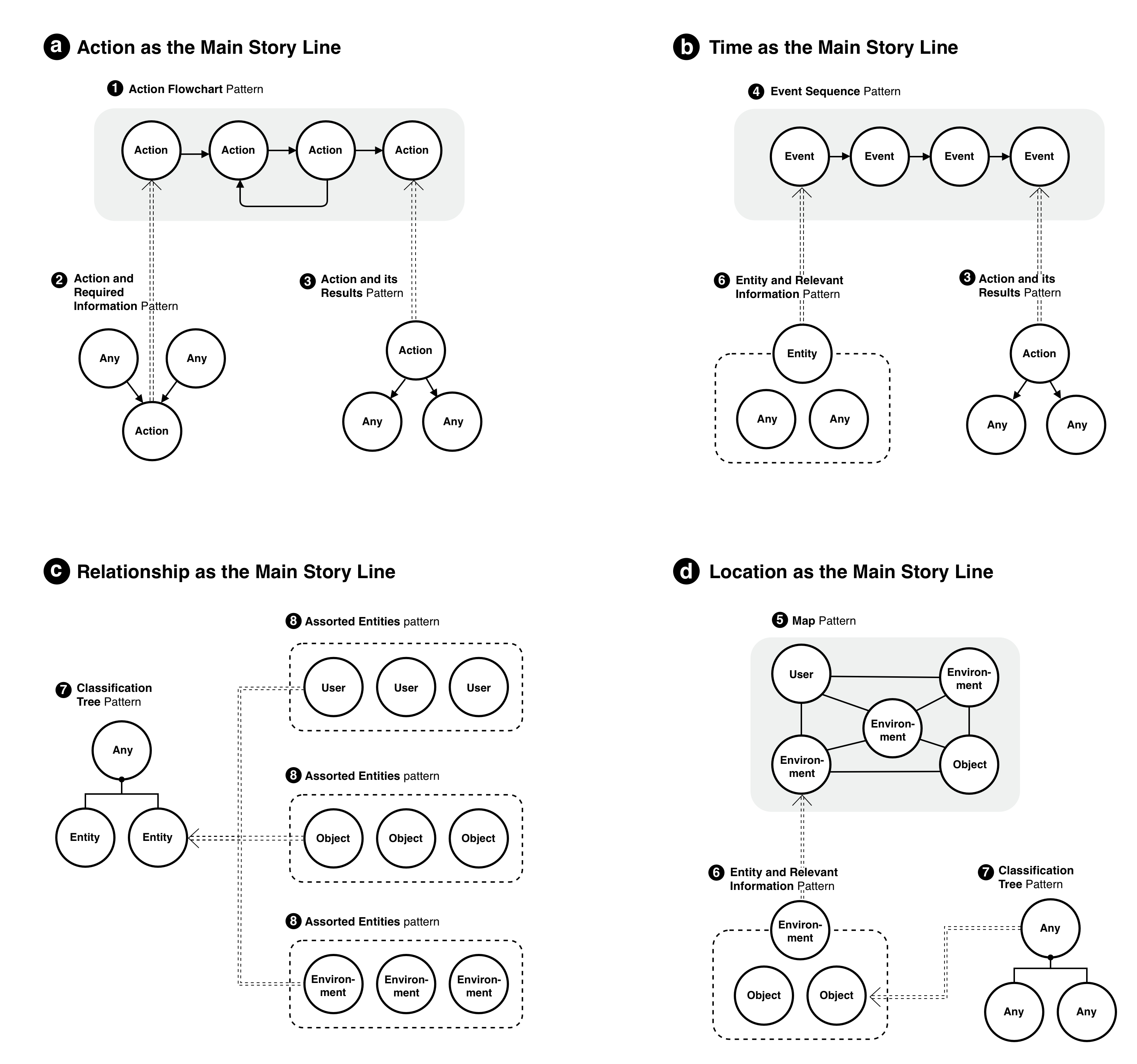}
  \caption{An overview of high-level patterns in our pattern library. Shaded parts indicate the ``main story line'' of the high-level pattern.}
  \Description{An overview of high-level patterns in our pattern library. Four patterns are present, each corresponding to a section below.}
  \label{fig:high-level-patterns}
\end{figure}

High-level patterns communicate a more complex message by combining and re-arranging low-level patterns. These patterns can be employed when designing interface modules for different knowledge types and scenarios. Figure~\ref{fig:high-level-patterns} presents an overview of the high-level patterns. Paragraphs below detailed each pattern with description, usage scenario, design considerations.

\hypertarget{high-pattern-1}{}
\subsubsection{Action as the Main Story Line}

\paragraph{Pattern description: } An \hyperlink{pattern-1}{\textit{Action Flowchart}} is the most prominent feature of this pattern, used often in combination with \hyperlink{pattern-2}{\textit{Action and Required Information}} or \hyperlink{pattern-3}{\textit{Action and Its Results}} to show information that's relevant to each action in the flowchart. 

\paragraph{When to use: } This pattern is often employed to explain the logical flow of robotic actions or to narrate the process of gathering specific information. Therefore, it mostly occurs in scenarios where the communicated knowledge is procedural or semantic.

\paragraph{Design considerations: } The action flowchart can be naturally represented on-screen as a flowchart, while relevant information is linked to the actions without affecting the flowchart structure. Participants used various low-level patterns to organize the related semantic information, such as \hyperlink{pattern-8}{\textit{Assorted Entities}} and \hyperlink{pattern-7}{\textit{Classification Tree}}. The link between the action and the other cards are depicted in various ways, including arrows, lines, or simply with proximity. This suggests that interface designers have a wide range of options when employing this pattern. 

In scenarios where mainly procedural knowledge is communicated, some of the participants neglected the semantic cards altogether and chose to present only the actions. Therefore, we suggest that the relevant information be collapsed into the \textit{action} element in a procedural scenario and remain hidden until the user specifically asks for details of the action.

\subsubsection{Time as the Main Story Line}

\paragraph{Pattern description: } This pattern features an \hyperlink{pattern-4}{\textit{Event Sequence}} as its main component. Other low-level patterns, especially \hyperlink{pattern-6}{\textit{Entity and Relevant Information}}, can be used to provide context.

\paragraph{When to use: } This pattern is most typically found in an episodic scenario, where the robot needs to show the knowledge related to a specific time in the past. 

\paragraph{Design considerations: } The \hyperlink{pattern-4}{\textit{Event Sequence}} can be naturally presented as a timeline. Both horizontal and vertical timelines have been observed during the study. However, no participants tried to show the exact time interval between two events on the timeline, even when such information was given. This suggests that when using this pattern, the ordering of events is essential while the exact timing of events is not. 
Many participants find the naming of the \textit{event} elements hard to understand, as they are from the robot's perspective (e.g. ``faces detected'', ``start interacting with user''). Some participants omitted events they deemed irrelevant, especially those without relevant semantic information. This suggests that designers should set rules to filter out irrelevant events to the user and translate the event names to user-friendly language when using this pattern.

\subsubsection{Relationship as the Main Story Line}

\paragraph{Pattern description: } 
This pattern features a large collection of semantic information, which is typically sorted into groups using \hyperlink{pattern-8}{\textit{Assorted Entities}}, or displayed as a \hyperlink{pattern-7}{\textit{Classification Tree}}. Sometimes a mix of the two is used, where part of the knowledge is organized in a tree-shaped structure and others categorized into groups.

\paragraph{When to use: } 
This pattern is often seen in semantic scenarios when the goal is to quickly convey a large amount of inter-related semantic knowledge to the user. A typical example is when the user wants to check the knowledge of a specific type. An interface that sorts knowledge by type can easily guide the user towards the goal.

\paragraph{Design considerations: } 
The \hyperlink{pattern-7}{\textit{Classification Tree}} can be shown as a collapsible tree. However, the designer may also choose to use hyperlinks between multiple screens to reduce clutter. Meanwhile, \hyperlink{pattern-8}{\textit{Assorted Entities}} can be represented most naturally as a list or group of elements on-screen. 

However, care must be taken when using \hyperlink{pattern-8}{\textit{Assorted Entities}}. A large number of elements of the same type can easily overwhelm the user. Many participants preferred \hyperlink{pattern-7}{\textit{Classification Tree}} and considered it a natural way to organize information. Some participants used symbols to represent cards frequently linked to and drew these symbols whenever a connection needs to be shown. This method can prove useful in the face of a large number of relations.

\subsubsection{Location as the Main Story Line}
\paragraph{Pattern description: } 
This pattern uses a \hyperlink{pattern-5}{\textit{Map}} as its core, linking other related knowledge to locations in the \hyperlink{pattern-5}{\textit{Map}}, using the \hyperlink{pattern-6}{\textit{Entity and Relevant Information}} pattern. 

\paragraph{When to use: } 
This pattern is typical in semantic scenarios and occurs when the knowledge to be communicated highly related to the location. Such knowledge includes objects, people, and named locations in space. However, participants also used this pattern when the robot's action involved traveling to multiple locations. 

\paragraph{Design considerations: } 
A map of the working environment would be a suitable background when using this pattern. The objects, users, or named locations can be placed on the map accordingly. When using this pattern to describe the robot's action, some participants used an abstract representation of the working environment, placing the locations in a sequence to show the robot's route. 

\section{Prototyping}

To ensure our pattern library's applicability, we further investigated how the patterns may be applied by prototyping a service robot. We conducted Wizard-of-Oz testing using our prototype and identified several design challenges and recommendations by analyzing the questionnaire and interview data. 

\subsection{Design}
We prototyped a service robot that took participants on a guided tour of the lab. The rationale behind choosing this scenario is two-fold. First, it is natural for the robot to communicate a large amount of knowledge of all three types (semantic, procedural, and episodic) during a tour. Second, it allowed us to apply real-world data to test the applicability of our patterns. 

Three researchers who were not involved in the pattern generation process were provided with the pattern library. They were given limited time (4 hours) to use the patterns to design a series of screens for displaying the necessary knowledge in a specific scenario. We didn't expect highly complete interfaces since the focus of the design session is to gain first-hand experience in applying the patterns.

\subsection{Implementation}

\begin{figure}[h]
  \centering
  \includegraphics[width=\linewidth]{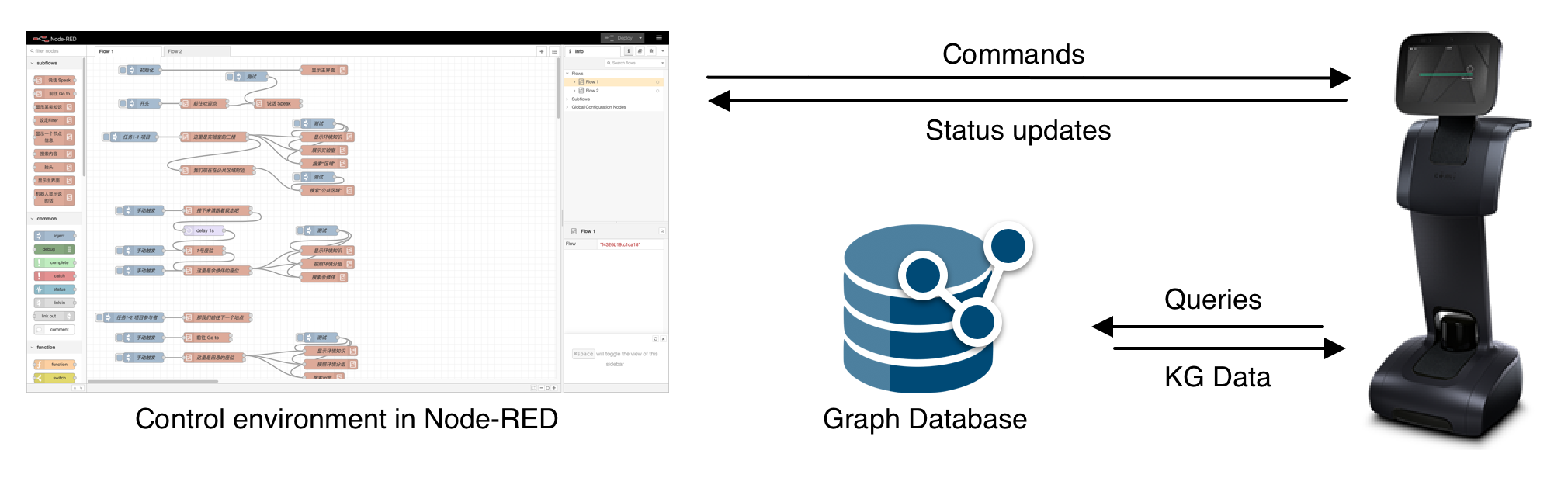}
  \caption{The Wizard-of-Oz testing system implemented in our study.}
  \Description{The Wizard-of-Oz testing system implemented in our study consists of a control environment in Node-RED, a graph database, and an Android application that runs on the Temi robot. The Temi application receives commands from Node-RED and queries information from the KG database.}
  \label{fig:wizard-of-oz-system}
\end{figure}

We implemented the Wizard-of-Oz testing system on the Temi robot\footnote{\url{https://www.robotemi.com}}. As shown in Figure~\ref{fig:wizard-of-oz-system}, our system consists of a graph database using Neo4j\footnote{\url{https://neo4j.com}} for querying the KG, a remote controlling environment using Node-RED\footnote{\url{http://nodered.org}}, and an Android application that runs on Temi that receives control commands and displays the UI\footnote{Available at \url{https://github.com/tongji-cdi/temi-woz-frontend} (dataset and UI) and \url{https://github.com/tongji-cdi/temi-woz-android} (Node-RED remote control and Temi Android application).}

We then compiled a knowledge graph dataset according to the types of situational knowledge (Table~\ref{tab:knowledgetype}). To test the applicability of our patterns to real-world problems, we curated data of the users, projects, objects, and locations from lab administration documents. We took care to anonymize the data by replacing person and object names.

\begin{figure*}[h]
  \centering
  \includegraphics[width=\linewidth]{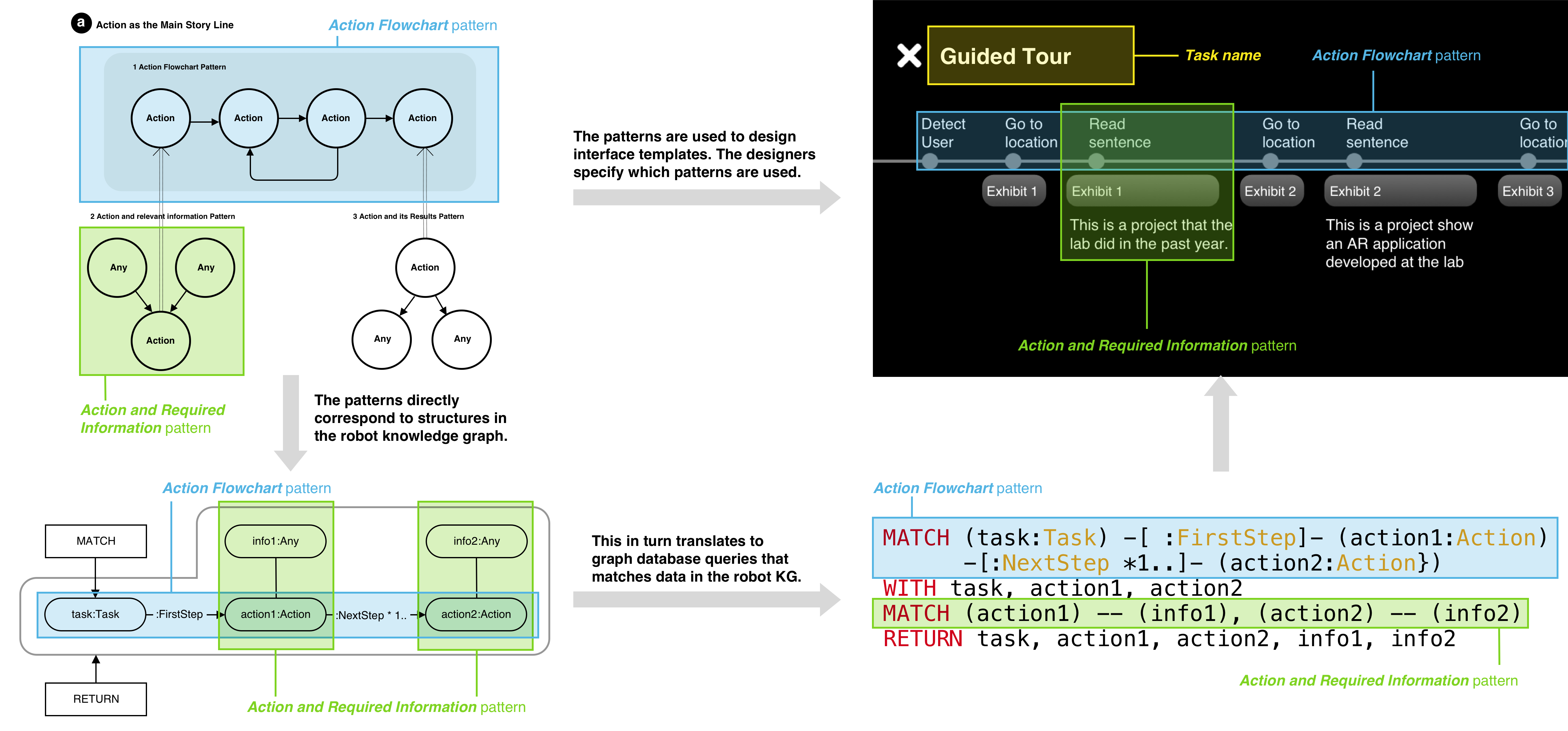}
  \caption{Turning patterns into interface design and a Cypher query.}
  \Description{The designer used the "Action as the Main Story Line" pattern, which consists of an action flowchart connected to the relevant information of the actions. The patterns are used to design an interface template. The designers specified which patterns are used. The patterns directly correspond to structures in the robot knowledge graph, which in turn translates to graph database queries that matches data in the robot KG. The match results are then used to fill information into  the user interface.}
  \label{fig:interface-pattern-query}
\end{figure*}

In the development process, our patterns demonstrated the ability to translate to knowledge graph queries quickly. An example of this is given in Figure~\ref{fig:interface-pattern-query}. The designer specified a task flowchart consisting of task steps and relevant KG nodes. The patterns were then directly translated to the Neo4j query language Cypher\footnote{\url{https://neo4j.com/developer/cypher}} to pull the knowledge from the database. 

\subsection{Testing}
Ten participants without expert knowledge of KGs performed in a simulated scenario with six tasks to complete. Participants assumed the role of a new research assistant and were guided by the robot to tour the lab. Along the tour, the accompanying researcher gave six tasks to the participant and recorded their completion status. Each task required the participant to look for some information using the interface on the robot's screen. The tasks cover the three scenarios (communicating semantic, procedural, and episodic knowledge) as outlined in Section \ref{section:scenarios}. Each scenario corresponds to two tasks, one requiring the participant to look for information on the current robot screen, and one requiring browsing the KG to search for the information.
 
 We designed a customized questionnaire based on the User Experience Questionnaire(UEQ) \cite{Schrepp_Eine} as few standard questionnaires target robotic interfaces aimed at communicating knowledge. The customized questionnaire contains four dimensions: usefulness, comprehensibility, perspicuity, and clarity. The questionnaire can be found in the supplementary materials.
 
 Interviews were conducted after participants filled out the questionnaires. We discussed the overall impression of the interface, the effects of the design, obstacles for the task, and their opinions on using screen-based interaction for situational knowledge exchange with the robot. The recordings of the interviews are transcribed for further analysis.

\subsection{Results}

\begin{figure}[h]
  \centering
  \includegraphics[width=\linewidth]{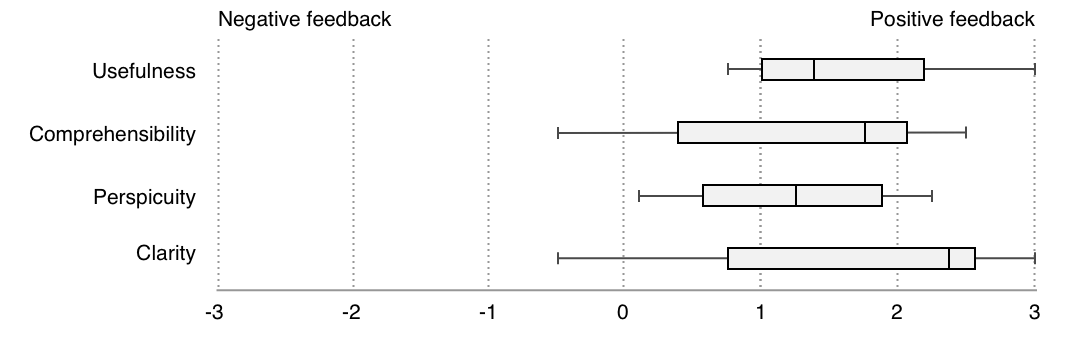}
  \caption{Questionnaire results from our Wizard-of-Oz testing. Scores range from -3 to 3, with positive numbers indicating a positive rating.}
  \Description{Boxplot of the questionnaire responses in Usefulness, comprehensibility, perspicuity, and clarity. Scores range from -3 to 3, with positive numbers indicating a positive rating. Mean scores are all above 1, while the mean score for Clarity is above 2. Scores for comprehensibility and perspicuity is slightly lower with smaller standard deviation.}
  \label{fig:questionnaire-score-overview}
\end{figure}

As seen in Figure~\ref{fig:questionnaire-score-overview}, participants gave an overall positive rating for all four aspects of the interface. This indicates that the patterns were able to support us in designing effective interfaces to communicate situational knowledge. Participants gave high ratings for the clarity of our interface (questions 13-16, mean=1.58, sd=1.69). Many consider this interface to be useful (question 1, mean = 1.90, sd = .88) and can provide help when using the robot (question 2, mean = 2.00, sd = .94).
Participants rate our interface as easy to learn (question 6, mean = 2.10, sd = .99). 

However, scores for comprehensibility and perspicuity is lower. Of note was that some participants gave lower scores for helpfulness in understanding the robot (question 7, mean=.70, sd=1.77). Some also think that the information presented can be confusing (question 12, mean=.80, sd=1.81). For some participants, it can be hard to find the information they want (question 11, mean=.80, sd=1.23).

The reasoning for the scores was discussed in the interviews that followed. In total, we recorded 122 minutes of interview audio. We analyzed the interview results while focusing on the successful and unsuccessful usage of patterns. We further discuss design recommendations generated from the analysis and our work's limitations in the following sections. 

\subsection{Design Recommendations}
Below, we highlight themes found in the interview data, and correlate design recommendations for the patterns in our library. 

\paragraph{1) Use user-friendly language when presenting KG data.}

All participants agreed that the classification of robotic knowledge into objects, users, and environment is intuitive. On the other hand, P10 felt uncertain what type of information to look for, and P3 proposed that the naming of knowledge types can be customized.

Therefore, we recommend that when using the \hyperlink{pattern-7}{\textit{Classification Tree}} pattern, naming the groups according to users' habits and customizing name by users can be considered. Moreover, it is advisable to translate robot ontology terms to more user-friendly names before display.

\paragraph{2) Avoid loops in the knowledge graph}

Repetitive information can confuse the user. P3 indicated that it feels confusing when entering a loop of often repeated information. They further commented on how it interfered with information retrieval. P8 thought some page jump is cyclic and repeated.
This is due to the inherent cyclic structure present in the KG. We recommend to remove cycles when applying the \hyperlink{pattern-5}{\textit{Map}} pattern, and organize information in a hierarchical fashion using \hyperlink{pattern-7}{\textit{Classification Tree}}.

\paragraph{3) Make it easier to look for information by relationships}

Four of the ten participants said the interface hierarchy was too deep. Three of them went back to the ontology list and looked through all nodes of a specific ontology type without following the relationships in the KG. Participants who found the information by following relationships felt it easier to complete the task than those who did not. Thus we recommend to use \hyperlink{pattern-6}{\textit{Entity and Relevant Information}} and enable the users to preview the information.

\paragraph{4) Use Classification Tree to reduce information overload.}

Users tend to reduce the complexity of information. P3 thought that the interface has a large amount of information and needed to spend more time to explore; P2 suggested to present less information at the beginning. P5 suggested that the information presented by the interface should be gradually increased according to the user's need. 

We have two design recommendations. First, \hyperlink{pattern-8}{\textit{Assorted Entities}} pattern is not applicable with a large amount of information at the same level. We recommend combine \hyperlink{pattern-8}{\textit{Assorted Entities}} to simplify information by common attributes or concepts. Second, using \hyperlink{pattern-7}{\textit{Classification Tree}} to present information from a large scale to a small scale is also recommended.

\section{Discussion and future work}
In this work, we described our process of developing a pattern library for designing interfaces to communicate the situational knowledge in service robots. Results from prototyping and analysis focused on three main findings: the use of design patterns, the challenges of presenting a knowledge graph to non-expert users, and human-robot interaction with a KG interface. These results offer many future directions for improving the pattern library and designing better interactions for human-robot knowledge exchange. 

\subsection{Design patterns}

During prototyping and testing, the pattern library was shown to aid our design and development process effectively. The patterns can be applied to real-world data, and interfaces designed using the patterns can easily use knowledge graph queries to populate its content. Our design session demonstrated that the patterns enable designers to produce interface designs without knowing the content of the robotic knowledge graph. 

Participants also responded favorably to the interface, highlighting its clarity in presenting robotic knowledge. Interviews with them provided us with insights on how to apply the patterns better. 

\subsection{Challenges of designing a KG interface}
Our testing also highlighted challenges that surface when designing a KG interface for non-expert users. We identified several challenges that originate from the knowledge graph structure, its content, and its size.

Participants often experienced confusion when loops occur in the KG structure. The patterns we discovered typically use categorization and trees to organize semantic information, which does not contain loops. This warrants further investigation into how to support efficient exploration of the KG while maintaining a hierarchical abstraction.

The naming of knowledge in the KG presents another challenge when displayed to non-expert users. The ontology of a KG is typically defined to facilitate effective programming without taking understandability into account. Work is being done to define an ontology that is understandable for non-expert users\cite{diprose_how_2012}. We believe that this is a necessary foundation that enables better human-robot knowledge exchange. 

The large amount of knowledge to be communicated poses another challenge. We have discussed how to apply hierarchical patterns to reduce clutter, but designers must also consider the increased depth of the interface.

\subsection{Designing better interactions to communicate situational knowledge}
Participants suggested various ways to incorporate multimodal interaction when using our interface. Some noted that using an interface is good when the robot is showing a large amount of information. Meanwhile, many participants also noted that it is more natural to use simple voice commands for querying a smaller amount of knowledge. This points toward a future research direction of using multimodal interaction (e.g., voice, gesture, and gaze) combined with a KG interface to enable better communication of situational knowledge.

\subsection{Limitations and Future Work}
The pattern library that we developed relied on our observations and analysis of twelve participants' knowledge presentation behavior under nine human-robot interaction scenarios. While we carefully mapped out these scenarios to cover as many potential cases as possible and recruited participants from diverse backgrounds, additional scenarios and testing may lead to the discovery of more patterns.

Moreover, the datasets we used are compiled specifically for the scenarios, which may have overlooked many complexities in real-world robot knowledge graphs. Future work may include building a public, real-world KG dataset. This will enable comparing the performance of patterns and interfaces, as well as promote further research into human-understandable robot knowledge ontology.

In the prototyping session, researchers with less practical knowledge in user interface design were tasked with designing the interfaces. While the participants gave positive feedback, results may be different had professional robot interaction and interface designers be employed to conduct the experiment. However, we hypothesize that this would likely improve participant feedback, which does not change our conclusions.

Finally, the pattern library we developed demonstrated its ability to apply to various knowledge communication scenarios. This warrants further research into developing authoring environments that support designers to design and prototype robot knowledge interfaces quickly.
\section{Conclusion}
This paper proposes a pattern library that shows how end-users envisioned service robots to organize and visually represent situational knowledge. We then prototyped a service robot based on the patterns and used Wizard-of-Oz testing to generate a series of design recommendations for knowledge-based interface design in robots. Future work would include using the patterns to create more robot interfaces and testing the design in more scenarios. We hope our work can inspire more researchers and interface designers to explore diverse approaches in situational knowledge exchange between humans and robots.

\bibliographystyle{ACM-Reference-Format}
\bibliography{main}

\end{document}